\documentclass[letterpaper]{article} 
\usepackage{aaai24}  
\usepackage{times}  
\usepackage{helvet}  
\usepackage{courier}  
\usepackage[hyphens]{url}  
\usepackage{graphicx} 
\usepackage{natbib}  
\usepackage{caption} 
\usepackage{pdfpages}
\usepackage{makecell}
\usepackage{multirow}
\usepackage{amssymb}

\usepackage{algorithm}
\usepackage{algorithmic}

\usepackage{amsmath}

\newcommand\modelname[1]{CrossBind}
\usepackage{newfloat}
\usepackage{listings}
\DeclareCaptionStyle{ruled}{labelfont=normalfont,labelsep=colon,strut=off} 
\floatstyle{ruled}
\newfloat{listing}{tb}{lst}{}
\floatname{listing}{Listing}
\title{CrossBind: Collaborative Cross-Modal Identification of \\Protein Nucleic-Acid-Binding Residues}
\author{
    Linglin Jing\textsuperscript{\rm 1,2}\footnote{These authors contributed equally. The work was done during their internships at the Shanghai Artificial Intelligence Laboratory.}, Sheng Xu\textsuperscript{\rm 3,1}$^{*}$, Yifan Wang\textsuperscript{\rm 2}, Yuzhe Zhou\textsuperscript{\rm 4}, Tao Shen\textsuperscript{\rm 3}, \\
    Zhigang Ji\textsuperscript{\rm 5}, Hui Fang\textsuperscript{\rm 2}, Zhen Li\textsuperscript{\rm 4}\footnote{Corresponding author.},
    Siqi Sun\textsuperscript{\rm 3,1 $\dagger$}\\
}
\affiliations{
    \textsuperscript{\rm 1}Shanghai Artifcial Intelligence Laboratory,\\

    \textsuperscript{\rm 2}Department of Computer Science, Loughborough University, 

     \textsuperscript{\rm 3}Research Institute of Intelligent Complex Systems, Fudan University, 
    
    \textsuperscript{\rm 4} SSE \& FNII, The Chinese University of Hong Kong (Shenzhen),

    \textsuperscript{\rm 5}Shanghai Jiao Tong University 

   \texttt{\small\{jinglinglin, xusheng1, sunsiqi1\}@pjlab.org.cn, lizhen@cuhk.edu.cn}

}

\usepackage{bibentry}

\begin{document}

\maketitle

\begin{abstract}
%
Accurate identification of protein nucleic-acid-binding residues poses a significant challenge with important implications for various biological processes and drug design.
Many typical computational methods for protein analysis rely on a single model that could ignore either the semantic context of the protein or the global 3D geometric information. Consequently, these approaches may result in incomplete or inaccurate protein analysis.
To address the above issue, in this paper, we present \modelname{1}, a novel collaborative cross-modal approach for identifying binding residues by exploiting both protein geometric structure and its sequence prior knowledge extracted from a large-scale protein language model. 
%
Specifically, our multi-modal approach leverages a contrastive learning technique and atom-wise attention to capture the positional relationships between atoms and residues, thereby incorporating fine-grained local geometric knowledge, for better binding residue prediction. 
%
%
%
Extensive experimental results demonstrate that our approach outperforms the next best state-of-the-art methods, GraphSite and GraphBind, on DNA and RNA datasets by {\textbf{10.8/17.3}}\% in terms of the harmonic mean of precision and recall (F1-Score) and {\textbf{11.9/24.8}}\% in Matthews correlation coefficient (MCC), respectively. We release the code at https://github.com/BEAM-Labs/CrossBind. 

\end{abstract}

\section{Introduction}
\label{sec:Intro}

Proteins and nucleic acids (DNA or RNA) interact in numerous biological processes, including regulation of gene expression, signal transduction, and post-transcriptional modification and regulation. Identifying protein nucleic-acid-binding residues with accuracy is critical for comprehending the mechanisms behind various biological activities and developing new drugs. However, direct measurement of protein binding sites is challenging and often not feasible, especially when large-scale analyses are conducted. This is because it requires time-consuming and expensive experimental techniques, such as X-ray crystallography or nuclear magnetic resonance spectroscopy. Therefore, computational prediction of binding residues in proteins with high efficiency and accuracy is essential.

\begin{figure}[!t]
\begin{center}
\includegraphics[width=0.8\linewidth]{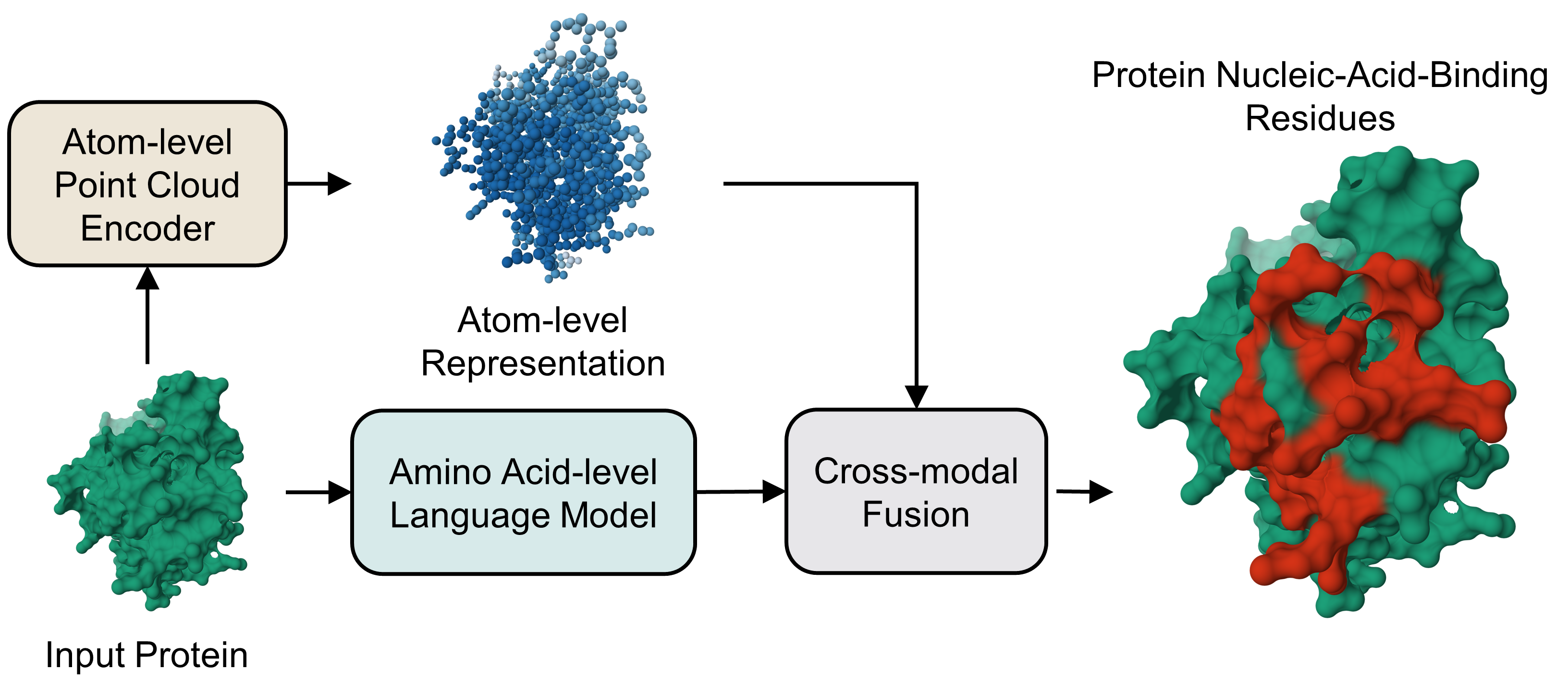}
\end{center}
   \caption{\modelname{1} model incorporates atom-level structure features in a point cloud representation and sequence features from a protein language model into a cross-model fusion module for protein Nucleic-Acid-Binding residues prediction (red).}
\label{fig:fig1}
\end{figure}

Proteins can be represented as strings of letters using the 20 distinct types of amino acids (AAs) and a residue refers to a specific AA in a protein chain. Atoms are the basic units of matter that make up everything in the universe, including AA. Several approaches have been developed for representing proteins computationally, including one-hot encodings~\cite{yan2016comprehensive}, position-specific scoring matrix (PSSM)\cite{su2019improving}, pseudo-AA composition\cite{chou2001prediction}, and hidden Markov models (HMM). Physico-chemical properties~\cite{chen2008predicting}, including hydrophobicity and electrostatics, have also proven effective for protein-related tasks. Currently, two primary types of protein-centric computational methods are available: sequence-based and structure-based methods.

Sequence-based methods analyze sequence-derived features to identify potential binding regions. Early machine learning methods for predicting binding residues were primarily based on the primary protein sequence~\cite{zhu2019dnapred,su2019improving,zhang2021ncbrpred}. However, their performance is limited because the patterns of binding residues are implicit in the spatial structure and cannot be identified from sequence information alone~\cite{wei2022protein}.

Structure-based methods use protein structures to identify binding residues and generally outperform sequence-based methods. 3D convolutional neural networks, graph neural networks, and their variants have been widely adopted in structure-based methods~\cite{lam2019deep,liu2013dnabind,xia2021graphbind}. However, structure-based methods typically require a large amount of biological information as training features, which consumes a lot of computing resources. Moreover, they may not accurately predict binding residues in cases where the protein or nucleic acid undergoes significant conformational changes upon binding~\cite{chen2021deep, dai2021protein}.

In recent years, the remarkable progress made in large-scale language modeling has extended to many fields, including the study of amino acids in proteins. Amino acids, which are characteristic of certain letters in proteins, have been the subject of study in recent works such as AlphaFold2~\cite{jumper2021highly}, which combines physical and biological knowledge about protein structure with deep learning algorithms implemented through transformer networks and 3D-equivariant structure transformation. Other works, such as RoseTTAFold~\cite{baek2021accurate}, ESMFold~\cite{verkuil2022language}, and ESM2~\cite{lin2022evolutionary}, have also been proposed, further improving the number of model parameters and computation efficiency. These developments are expected to have a significant impact on downstream protein function studies, such as the prediction of residues.

To address the limitations of the single-mode method, we present a new cross-modal training approach, named \modelname{1}, for identifying nucleic-acid-binding residues using both protein structure and sequence information. The proposed method leverages the power of deep learning to facilitate interactions between structure and sequence features at multiple scales, resulting in improved cross-modal fusion and utilization. The overview architecture of \modelname{1} is illustrated in Figure~\ref{fig:overview}. The sequence encoder component employs ESM-2~\cite{lin2022evolutionary}, one of the largest protein language models to date, which was trained on millions of protein sequences with 15B model parameters. The structure encoder, on the other hand, uses a sparse convolution encoder~\cite{schmohl2019submanifold} to represent residues as a point cloud segmentation task at the atom-level. To capture the positional relationships between atoms and residues, an atom-wise attention (AWA) mechanism is introduced since the interactions between proteins and nucleic acids can occur on both backbone and side-chain atoms. Additionally, we introduce a self-supervised learning (SSL) strategy to account for conformational changes in 3D protein structures, increasing the diverse mobility of atoms and enhancing their ability to transmit signals when interacting with other molecules. Furthermore, since our dataset is imbalanced, we employ SSL to enhance the robustness of our model~\cite{liu2021self}.

In summary, our main contributions are listed as follows:
\begin{itemize}
\item We propose a novel cross-modal strategy, \modelname{1}, that combines protein structure and sequence information to identify nucleic-acid-binding residues. 

\item Our method employs an atom-level point cloud segmentation on residues, along with an atom-wise attention component, to efficiently extract fine-grained local geometric knowledge of protein structure. 

\item We incorporate several biological task-related modules and demonstrate that our approach achieves state-of-the-art performance on multiple datasets consistently.
\end{itemize}

\section{Related work}
\label{sec:RW}

\subsection{Sequence-based method}

Sequence-based methods offer a flexible approach to predicting protein-nucleic-acid-binding residues that can be applied to any protein sequence. There are two main types of sequence-based models: alignment-based methods and machine-learning-based models. Alignment-based methods rely on the assumption that proteins with similar sequences share similar binding partners and binding residues~\cite{xue2011homppi}. These methods predict binding residues by comparing annotations from proteins in the database that are sufficiently similar to the input protein. To do this, they typically require a database containing annotations of known binding residues. Sequence similarity between protein chain pairs can be calculated using E-value~\cite{mcginnis2004blast} and TM-align~\cite{zhang2005tm}. Machine-learning-based methods, on the other hand, predict the probability of each residue binding or not by leveraging sequence contextual information. Each protein residue is encoded with a feature vector as the model input, which typically contains physicochemical characteristics of the predicted residue and its neighboring residues. Examples of such methods~\cite{ pan2018predicting, gronning2020deepclip} typically employ 1D convolution layers and bidirectional LSTM to capture local and global features from the protein sequence for binding prediction.

Recent progress in protein language models, such as ESM2, has enabled the utilization of pre-trained models for processing protein sequences. Leveraging the vast amount of data regarding the physical and chemical properties of protein structures, these models have displayed remarkable accuracy in predicting protein structures and executing a wide range of downstream tasks based exclusively on protein sequences~\cite{lin2022evolutionary}.

\begin{figure*}[!t] 
\begin{center}
\includegraphics[width=0.8\linewidth]{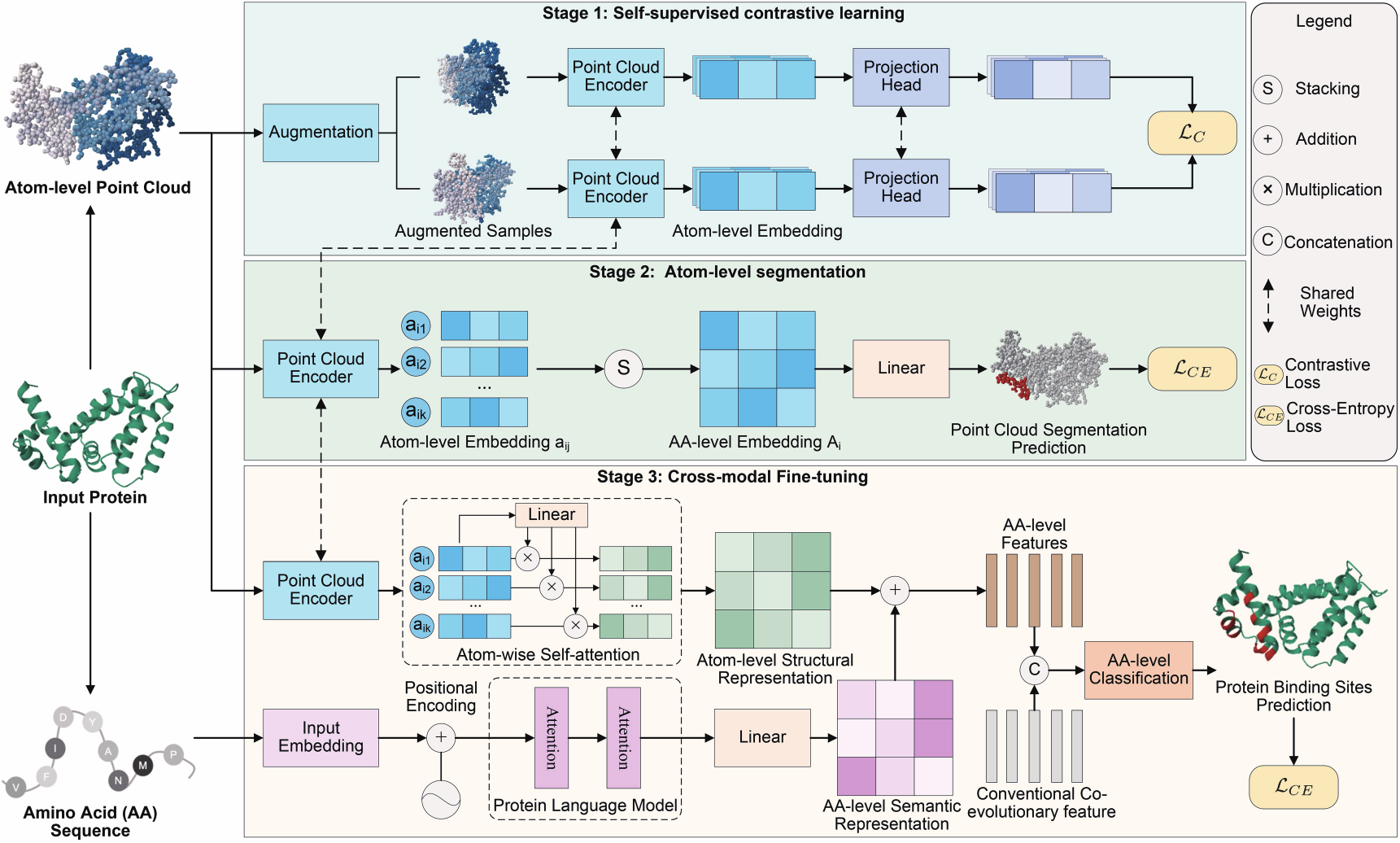}
\end{center}
\caption{The overall architecture of \modelname{1}. Given a query protein, the input comprising atom-level structure information is fed to the Point Cloud Encoder. The encoder, which is pre-trained by using a self-supervised learning strategy, generates a structural point cloud representation. Further, an atom-wise attention module is introduced to capture the positional relationships between atoms and residues. Finally, the cross-modal module combines the structural and sequence representations to concatenate with co-evolutionary features for the prediction of protein binding sites.}
\label{fig:overview}
\end{figure*}

\subsection{Structure-based method}

Recent structure-based methods~\cite{lam2019deep,xia2021graphbind} use low-resolution structural information, such as spatial neighbors, solvent accessibility, and secondary structure~\cite{liu2013dnabind}, derived from protein structures to predict binding residues. These methods employ different approaches, such as constructing graphs or 3D-CNNs as spatial representation encoders. GraphBind~\cite{xia2021graphbind}, for example, proposes a hierarchical graph neural network that learns protein structural context embeddings for recognizing nucleic-acid-binding residues by using the residue and its physicochemical properties as nodes and the positional distance between residues as edges to construct a graph. However, due to the non-Euclidean nature of protein structures, learning latent knowledge from structures remains one of the most significant challenges~\cite{wei2022protein, bheemireddy2022computational}.

 On small-molecule-level tasks, such as ligand prediction, point cloud-based encoders are widely used~\cite{yan2022pointsite,wang2022point}. Due to the distribution state of small molecules (atoms) in protein, they can be expressed as a form of the point cloud. However, due to the complexity of residues, which contain multiple atoms, representing them in point cloud form presents a challenge. While deep learning methods based on point clouds as high-resolution structure encoders have been successful in computer vision and autonomous driving~\cite{guo2020deep}, applying these methods to protein residues remains an active area of research.

\subsection{Cross-modal learning}
Learning from multiple modalities can provide rich learning signals that enable the extraction of semantic information from a given context~\cite{zhou2019review}. Recent studies have shown that cross-modal learning, which combines information from different modalities, can achieve better results than using a single modality alone~\cite{ding2021vision,panda2021adamml}. The use of cross-modal learning has many potential applications, including in the fields of computer vision, natural language processing, and robotics, where it can help to improve the accuracy and efficiency of tasks such as object recognition, speech recognition, and machine translation.

Protein structure and sequence information can be seen as two distinct modalities that provide complementary information for predicting protein properties. Graphsite~\cite{yuan2022alphafold2} introduced a transformer that combines sequence and structure information to predict DNA-binding residues. Specifically, Graphsite uses AlphaFold2 to represent the sequence and maps the protein residues to a distance matrix that captures the pairwise distance relationship between each residue. However, using only the distance matrix to represent the 3D structure may result in a loss of spatial information, which is not considered true cross-modal learning.

\section{Methods}
\label{sec:MD}
This paper presents a novel cross-modal learning framework, called \modelname{1}, that aims to enhance the identification of protein nucleic-acid-binding residues. The approach leverages both atom point clouds and amino acid sequences to learn a unified representation of protein structure and sequence. The paper is organized as follows: first, we introduce the pre-training of the atom point cloud segmentation in Section~\ref{subsec:ALS}. Next, we present the details of our cross-modal that integrates information from both protein structure and sequence in Section~\ref{subsec:Multi}. Finally, we describe a filter module in Section~\ref{subsec:filter} that is designed to leverage protein structure and biological properties and achieve further performance improvements. The overall approach is illustrated in Figure~\ref{fig:overview}.


\subsection{Definitions and problem formulation}
\label{subsec:Definitions}


Our goal is to identify binding residues in a given query protein, using a cross-modal training strategy that combines atom-level point cloud segmentation (ALS) with a protein large language model (LLM). The input data for each protein consists of both structural and sequence information. The structural data is composed of atom point clouds, each with three spatial coordinates $(X, Y, Z)$, while the sequence data consists of amino acids with varying lengths ranging from tens to thousands. There are 21 AA types and 5 atom types in each protein.


To learn a segmentation encoder that can effectively represent protein 3D structure information, we employ a self-supervised learning strategy using the sparse convolution encoder $F_\Theta$ and multi-layer perceptron (MLP) projection heads $G_\psi$ on unlabeled atom cloud points. The learned representations are then used for downstream tasks. After applying $F_\Theta$, each atom is represented by a 32-dimensional vector $a_{ij}\in \mathbb{R}^{32}$. To obtain an amino acid level representation $A_{i}\in \mathbb{R}^{448}$, we stack the atom representations $a_{ij}$ corresponding to the $j$-th atoms in the $i$-th amino acid.


In the cross-modal module, we employ an atom-wise attention (AWA) mechanism to process the atom-level representations $\left\{a_{i 1}, a_{i 2}, \ldots, a_{i k}\right\}$ obtained from the $F_\Theta$, where $k$ denotes the maximum index of atom in amino acid. This mechanism captures the positional relationships between the atoms and residues and produces a new representation $F_{struct}\in \mathbb{R}^{L\times448}$ that encodes the structural information, where $L$ denotes the number of AAs in a protein. Simultaneously, the sequence information is captured by a large language model (LLM), which produces a sequence representation $F_{seq} \in \mathbb{R}^{L\times1280}$. In addition, we consider the conventional biological co-evolutionary feature $F_{evo} \in \mathbb{R}^{L\times54}$ as an extra feature for identifying nucleic-acid-binding residues. The details of how these cross-modal features are fused are described in Section~\ref{subsec:Multi}.

\subsection{Atom-level segmentation (ALS) module}
\label{subsec:ALS}

{\bf Point cloud encoder.} For protein structure analysis, we adopt a sparse convolution-based U-net~\cite{schmohl2019submanifold} as the segmentation encoder $F_\Theta$. To accomplish this, we convert the original protein atoms into point clouds that include atom coordinates and features. These features are one-hot embeddings of amino acids and atoms. Consequently, the input for segmentation encoder includes 27-dimensional atom features and 3-dimensional spatial coordinates. The ALS output, $a_{ij} \in \mathbb{R}^{32}$, provides atom-level spatial information. However, binding sites identification occurs at the amino acid level; Hence, we employ a padding strategy to stack $a_{ij}$ onto the amino acid level $A_i$: 

\begin{equation}
A_i= \left[\hat{a}_{i 1}, \hat{a}_{i 2}, \ldots, \hat{a}_{i K}\right], \hat{a}_{ij} =  \begin{cases}a_{ij} & j \leq k \\ \textbf{0} & k < j \leq K\end{cases}
\label{eq:a2A}
\end{equation}

where $i$ denotes the index of an amino acid, $j$ denotes the index of an atom within an amino acid and $k$ represents the maximum number of atoms within any given amino acid. $[.]$ denotes the concatenate function, and $K$ is a constant that serves as an upper limit on the number of atoms any amino acid can contain in the entire data set.

{\noindent\bf Self-supervised learning (SSL).} To address the issue of atom mobility in protein 3D structures, we use SSL to improve the identification of conformational changes that occur during binding. Furthermore, the success of self-supervised learning in handling imbalanced data has motivated us to use it to enforce invariance to a set of point cloud geometric transformations. Our approach involves using protein atoms point clouds as input and constructing augmented versions ${Q}^{t_1}$ and ${Q}^{t_2}$ using randomly combined transformations that include normal transformations such as rotation, scaling, and translation, as well as spatial transformations such as elastic distortion and jittering. The point cloud encoder $F_\Theta$ maps atom point clouds to the feature embedding space, and the feature embedding is then projected to an invariant space with projection heads $G_\psi$. We denote the projected vectors as $z_i^{t1}$ and $z_i^{t2}$, where $z_i^t = F_\Theta(G_\psi({Q_i}^{t}))$. Similar to SimCLR~\cite{chen2020improved}, we use the NT-Xent loss as the contrastive loss (Section~\ref{subsec:Loss}).

\subsection{Cross-modal module}
\label{subsec:Multi}

{\noindent\bf Atom-wise attention (AWA).} The structure of our AWA module, which is illustrated in Figure~\ref{fig:overview}. The AWA module is designed to dynamically highlight the centroid of each residue, encompassing both backbone and side-chain atoms, and generate the residue representation by stacking its constituent atoms. The atom representations $a_{ij}$ from a residue are combined to form the global representation between all atoms. The atom-wise attention score $\Lambda=\left\{\left.\sigma_j\right|j=1,2, \ldots K\right\}$  is calculated using a simple MLP mechanism~\cite{chen2022part}, and a $Sigmoid$ function is used to map the attention value to a range of (0, 1).

\begin{equation}
\Lambda=\operatorname{Sigmoid}\left(\operatorname{MLP}\left(\left[a_{i 1}, a_{i 2}, \ldots, a_{i K}\right]\right)\right),
\label{eq:attention}
\end{equation}
%
According to Eq~\ref{eq:attention}, each element in $\Lambda$ indicates the importance of an atom in the same residue. The attention score is used to weight the atom representations $a_{ij}$ by element-wise multiplication. Finally, we concatenate the weighted atom representations and use an MLP layer to generate the protein structure representation $F_{struct}$.
\begin{equation}
F_{struct}=\operatorname{MLP}\left([{a_{i1} \odot \sigma_1,a_{i2} \odot \sigma_2,...,{a_{iK}} \odot \sigma_K}]\right)),
\end{equation}
where $K$ has the same definition as Eq~\ref{eq:a2A}, $\sigma_K$ is the K-th scale element from $\Lambda$, $\odot$ denotes the element-wise multiplication to scale weight the atom point cloud representations.

{\noindent\bf Cross-modal fusion.} We used three groups of protein features to train our model: protein structural representation $F_{struct}$, sequence representation $F_{seq}$, and conventional biological co-evolutionary feature $F_{evo}$. $F_{seq} \in \mathbb{R}^{L \times 1280}$ first compress the feature dimension to the same as $F_{struct} \in \mathbb{R}^{L \times 448}$ by MLP, where $L$ denotes the number of amino acids in a protein. Then $F_{seq}$ and $F_{struct}$ were fused together with the following fusion rules: 

\begin{equation}
P_o=\lambda F_{struct}+(1-\lambda) F_{seq},
\label{eq:fusion}
\end{equation}
where $\lambda$ denotes a learnable parameter, '$+$' denotes the element-wise addition.
At last, conventional biological co-evolutionary
feature $F_{evo} \in \mathbb{R}^{L \times 54}$ was concatenated with $P_o$ as the final group feature $P_{final} \in \mathbb{R}^{L \times 502}$ for the binding residue identification:

\begin{equation}
P_{final}=\left(\left[P_o, F_{evo}\right]\right).
\label{eq:final_out}
\end{equation}

\subsection{Residue propensity filter (RPF)}
\label{subsec:filter}
As highlighted earlier, traditional approaches for identifying binding sites often require additional information, such as geometric or charge distribution, to accurately define binding regions. However, our proposed \modelname{1} method utilizes the inherent characteristics of proteins to identify binding residues. Specifically, binding residues are known to predominantly occur on the surface of proteins, which inspired us to design a biological filter that enhances the interpretability of the identification task. We adopted the amino acid propensity~\cite{kim2006amino}, which measures the likelihood of an amino acid to interact with nucleic-acids, as a filtering condition. This biological property, such as the higher propensity of positively charged amino acids to interact with nucleic-acids, has been well established through biological experiments. Additionally, we leveraged the Geodesic-distance~\cite{sverrisson2021fast}, a measure of the distance between amino acids on the protein surface, to screen for outliers within a certain range. The outlier amino acids were then used to scale the logits of the \modelname{1} output according to their corresponding propensity, thereby improving the accuracy of the model. The amino acid propensity is calculated as follows:

\begin{equation}
\xi^i= (\frac{\bar{n}^i}{\sum_{i=1}^{20} \bar{n}^i}) / (\frac{{n}^i}{\sum_{i=1}^{20} {n}^i}),
\end{equation}

where ${n}_i$ denotes the number of amino acid $i$ with label 0 and ${\bar{n}_i}$ is label 1. $\xi$ is calculated on the training set.

The algorithmic specifications of the RBF method can be found in the Supplementary Material.



\subsection{Loss functions}
\label{subsec:Loss}
{\noindent\bf Classification loss.} Given a training set $V_{t r}$, in ALS and Cross-Modal module, we use cross-entropy loss as:

\begin{equation}
\mathcal{L}={-}\sum_{V_{t r}}\left(y_i \ln \widehat{y}_i+\left(1-y_i\right) \ln \left(1-\widehat{y_i}\right)\right),
\end{equation}
where $y_i$ is the label of a residue and $\widehat{y}_i$ is the probability corresponding to $y_i$. 

{\noindent\bf Contrastive loss.} In SSL strategy, we leverage NT-Xent loss, which maximize the similarity of $(\mathbf{z}_i^{t_1}$,$\mathbf{z}_i^{t_2})$ and minimizing the similarity with all the other samples in the mini-batch of point clouds. The loss function for a positive pair
of examples $(i, j)$ is defined as:

\begin{equation}
S_{i,i}^{t1,t2}=\cos \operatorname{sim}(z_i^{t1},z_i^{t2}),
\label{eq:loss_C}
\end{equation}

\begin{small}
\begin{equation}
\mathcal{L}(i)=-\log \frac{\exp \left(S_{i,i}^{t_1,t_2} / \tau\right)}{\sum_{\substack{j=1 \\ j \neq i}}^N \exp \left(S_{i,j}^{t_1,t_1}) / \tau\right)+\sum_{j=1}^N \exp \left(S_{i,j}^{t_1,t_2} / \tau\right)},
\end{equation}
\end{small}
where $\cos \operatorname{sim}(\cdot)$ denotes the cosine similarity function. $N$ is the mini-batch size, $\tau$ is a pre-set temperature constant. $z_i^t$ denotes the projected vector.


\section{Experiments}
\label{sec:exp}

\subsection{Experiments Setup}
\label{subsec:Setup}
{\noindent\bf Datasets and evaluation metrics.} We utilized two benchmark datasets, DNA\_129 and RNA\_117 dataset, from a previous study for training and testing our method. These datasets were obtained from the BioLip database~\cite{yang2012biolip} and consist of experimentally determined complex structures. The DNA\_573 comprises 573 training proteins and 129 testing proteins, while the RNA\_117 dataset consists of 495 training proteins and 117 testing proteins. A binding residue was identified if the smallest atomic distance between the target residue and the DNA molecule was less than 0.5 Å plus the sum of the Van der Waal’s radius of the two nearest atoms. To demonstrate the generalizability of our method, we used an additional independent testing set: DNA\_181, which contained 181 proteins whose structures were predicted by Alphafold2. The datasets are highly imbalanced, with a large ratio between positive and negative samples.

To evaluate the performance of our method, we used several commonly metrics, including precision (Pre), recall (Rec), F1-score (F1), Matthews correlation coefficient (MCC), area under the receiver operating characteristic curve (AUC), and area under the precision-recall curve (AUPR). AUC and AUPR are threshold-independent measures, providing an overall assessment of the model's performance. The remaining metrics require the use of a threshold to convert predicted binding probabilities into binary predictions, which we determined by maximizing the F1-score. As the test set is highly imbalanced, F1, MCC, AUC, and AUPR are more suitable as overall metrics. All reported metrics are the averages of ten repeated runs of the method. The performance metrics are summarized in Table~\ref{table:performance_sum}.


{\noindent\bf Implementation details.} In all experiments, we used the Adam optimizer with a weight decay of $1 \times 10^{-4}$ and employed cosine annealing as the learning rate scheduler. We set the initial learning rate to $1 \times 10^{-3}$ for the ALS module and $1 \times 10^{-4}$ for the cross-modal module. In the RPF module, we ranked the nearest neighbors and selected the top five amino acids for the propensity filter using prediction logits with thresholds of [-0.8, 0.8], corresponding to a positive over a negative propensity. For data processing, we centered and re-scaled each point cloud to fit into a sphere, and then represented it as a sparse voxel representation with 0.1 voxel size. We trained and validated our method on the training dataset, with a validation ratio of 0.1.

\begin{figure}[!b]
\centering
\footnotesize
\includegraphics[width=1\linewidth]{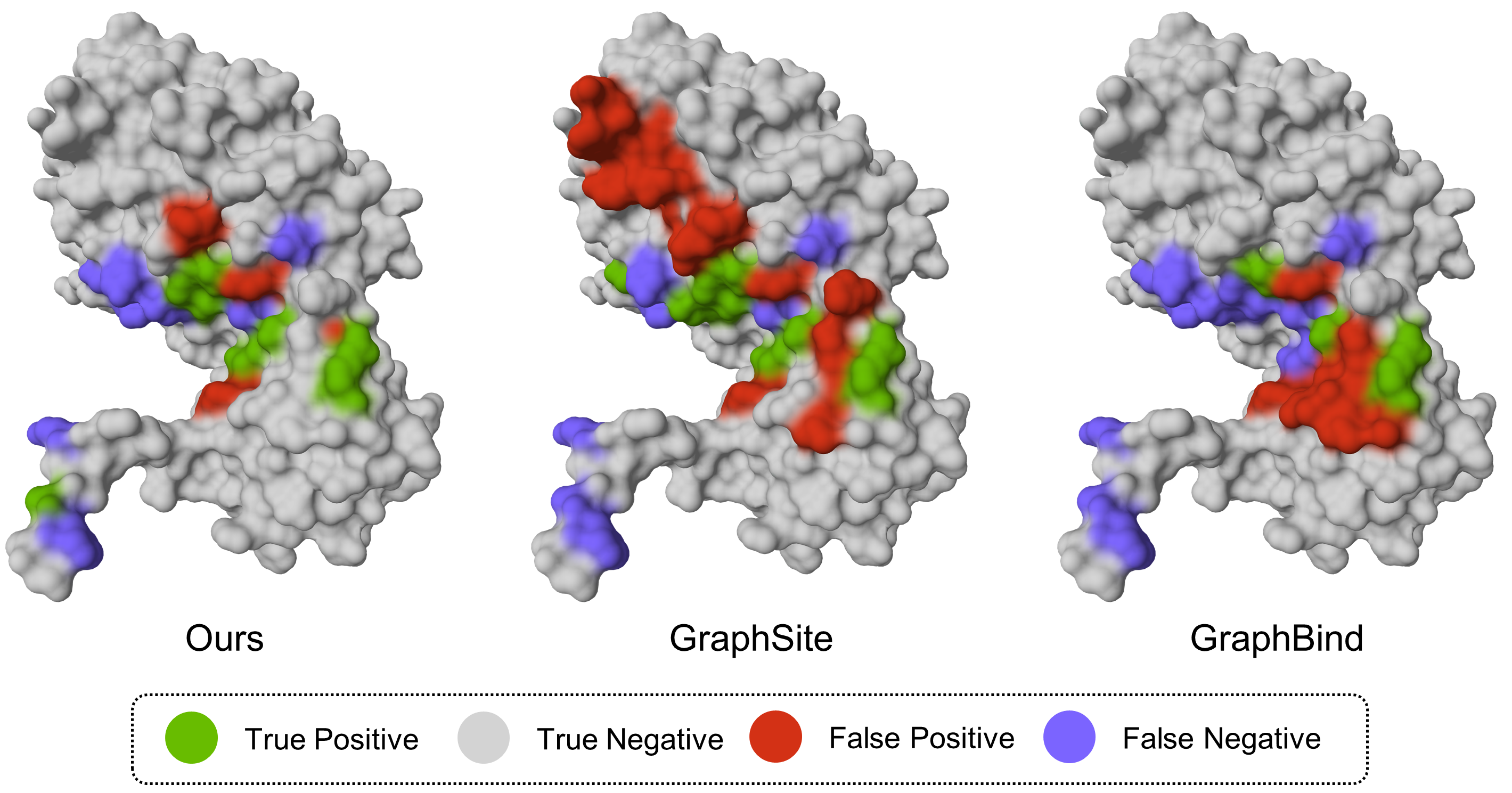}
   \caption{Classification results for protein chain 6YMW\_B using \modelname{1}(Ours), GraphSite and GraphBind.}
\label{fig:case}
\end{figure}


\begin{table*}[t]
\centering
\footnotesize
\begin{tabular*}{1\textwidth}{@{\extracolsep{\fill}}lcccc|cccccc} 
\hline
Dataset & Method & Struct & Seq & LLM  & Rec & Pre & F1 & MCC & AUC & AUPR  \\
\hline

\text { DNA\_129  } & COACH-D~\cite{wu2018coach} & $\checkmark$ & & & 0.328 & 0.318 & 0.323 & 0.279 & 0.712 & 0.248  \\
                        & NucBind~\cite{su2019improving} & $\checkmark$ && & 0.322 & 0.366 & 0.343 & 0.304 & 0.809 & 0.284  \\
                    & SVMnuc~\cite{su2019improving} &  &$\checkmark$ & & 0.316 & 0.371 & 0.341 & 0.304 & 0.812 & 0.302  \\
                     & NCBRPred~\cite{zhang2021ncbrpred} &  &$\checkmark$& & 0.312 & 0.392 & 0.347 & 0.313 & 0.823 & 0.310  \\
                     & DNABind~\cite{liu2013dnabind} &$\checkmark$ && & 0.487 & 0.389 & 0.433 & 0.395 & 0.832 & 0.391  \\
                     & DNAPred~\cite{zhu2019dnapred} & &$\checkmark$& & 0.396 & 0.353 & 0.373 & 0.332 & 0.845 & 0.367  \\
                    & GraphBind~\cite{xia2021graphbind} & $\checkmark$&&  & 0.625 & 0.434 & 0.512 & 0.484 & 0.916 & 0.497  \\  
                    & GraphSite$^\text{a}$~\cite{yuan2022alphafold2} &&$\checkmark$&$\checkmark$ & 0.665 & 0.460 & 0.543 & 0.519 & 0.934 & 0.544  \\ 
                    \hline
                        & {\bf \modelname{1}} & $\checkmark$&$\checkmark$&$\checkmark$  & {\bf 0.684} & {\bf 0.538} & {\bf 0.602} & {\bf 0.581} & {\bf 0.953} & {\bf 0.628}  \\ 
                    \hline
\text { DNA\_181  } & COACH-D~\cite{wu2018coach}& $\checkmark$ & && 0.254 & 0.280 & 0.266 & 0.235 & 0.655 & 0.172  \\
                     & NCBRPred~\cite{zhang2021ncbrpred} & $\checkmark$ && & 0.259 & 0.241 & 0.250 & 0.215 & 0.771 & 0.183  \\
                    & SVMnuc~\cite{su2019improving} &  &$\checkmark$& & 0.289 & 0.242 & 0.263 & 0.229 & 0.803 & 0.193 \\ 
                     & NucBind~\cite{su2019improving} & &$\checkmark$& & 0.293 & 0.248 & 0.269 & 0.234 & 0.796 & 0.191  \\
                    & DNABind~\cite{liu2013dnabind} & $\checkmark$ && & 0.535 & 0.199 & 0.290 & 0.279 & 0.825 & 0.219  \\
                    & DNAPred~\cite{zhu2019dnapred} & &$\checkmark$ & & 0.334 & 0.223 & 0.267 & 0.233 & 0.655 & 0.172 \\
                    & GraphBind~\cite{xia2021graphbind} & $\checkmark$ & & & 0.505 & 0.304 & 0.380 & 0.357 & 0.893 & 0.317 \\
                    & GraphSite$^\text{a}$~\cite{yuan2022alphafold2} &&$\checkmark$&$\checkmark$ & 0.517 & 0.354 & 0.420 & 0.397 & 0.917 & 0.369  \\ 
                    \hline
                    & {\bf \modelname{1}} & $\checkmark$&$\checkmark$&$\checkmark$  & {\bf 0.538} & {\bf 0.432} & {\bf 0.475} & {\bf 0.448} & {\bf 0.932} & {\bf 0.424}  \\ \hline
 \text { RNA\_117  } & RNABindR Plus~\cite{yu2013designing} & &$\checkmark$& & 0.273 & 0.227 & 0.248 & 0.202 & 0.717 & -  \\        
                         & SVMnuc~\cite{su2019improving} & &$\checkmark$& & 0.231 & 0.240 & 0.235 & 0.192 & 0.729 & -  \\
                         & COACH-D~\cite{wu2018coach} & $\checkmark$ && & 0.221 & 0.252 & 0.235 & 0.195 & 0.663 & -  \\
                         & NucBind~\cite{su2019improving} & $\checkmark$ && & 0.231 & 0.235 & 0.233 & 0.189 & 0.715 & -  \\
                         & aaRNA~\cite{li2014quantifying} & $\checkmark$&& & 0.484 & 0.166 & 0.247 & 0.214 & 0.771 & -  \\
                         & NucleicNet~\cite{lam2019deep} & $\checkmark$ & && 0.371 & 0.201 & 0.261 & 0.216 & 0.788 & -  \\
                         & GraphBind~\cite{xia2021graphbind} & $\checkmark$ && & 0.463 & 0.294 & 0.358 & 0.322 & 0.854 & -  \\
                         \hline
                        & {\bf \modelname{1}} & $\checkmark$&$\checkmark$&$\checkmark$ & {\bf 0.490} & {\bf 0.366} & {\bf 0.420} & {\bf 0.402} & {\bf 0.903} & {\bf 0.352}  \\ 
    
\hline
\end{tabular*}
\caption{Performance comparison of \modelname{1} with state-of-the-art methods on nucleic-acid-binding tasks. Structure\_based and Sequence\_based method are listed in the table as two main training methods, and protein large language model (LLM) is also used in some work to improve the performance. $^a$ Using the predicted structure by AlphaFold2.}
\label{table:performance_sum}
\end{table*}

\subsection{Comparison with state-of-the-art methods}
\label{subsec:Compare_result1}

We evaluated the performance of our method against state-of-the-art methods on three nucleic-acid-binding test sets: DNA\_129, DNA\_181, and RNA\_117, as reported in previous works~\cite{xia2021graphbind,yuan2022alphafold2}. The previous methods contained both protein structure-based and sequence-based methods and are shown in Table~\ref{table:performance_sum}. Similar to GraphSite, we used the large language model as the protein sequence encoder. As shown in Table~\ref{table:performance_sum}, our method outperformed the second-best sequence-based method, GraphSite, \modelname{1} improved F1-score, MCC, and AUPR by 10.8\%, 11.9\%, and 15.4\%, respectively. \modelname{1} demonstrated superior predictive accuracy, outperforming the 
second-best structure-based method, GraphBind by 17.5\%, 20.0\%, and 26.3\% in F1-score, MCC, and AUPR, respectively. To demonstrate the generalization and stability of our method, we also compared \modelname{1} with other methods on two more challenging test sets: DNA\_181 and RNA\_117. The performance ranks of these methods are generally consistent with those in Test\_129, and \modelname{1} still outperforms all other methods significantly. When using GraphSite as the baseline on the DNA\_181 dataset, \modelname{1} improved F1-score, MCC, and AUPR by 13.0\%, 12.8\%, and 14.9\%, respectively. When using GraphBind as the baseline on the RNA\_117 dataset, \modelname{1} improved F1-score and MCC by 17.3\% and 24.8\%, respectively.

{\noindent\bf Case studies.} To analyze our results in more detail, we conducted a visualization of three cases predicted by GraphSite, GraphBind, and our proposed model on DNA\_129. We selected the example protein 6YMW\_B, which was also discussed in~\cite{yuan2022alphafold2}. This protein contains 668 residues, out of which 13 are binding residues. As demonstrated in Figure~\ref{fig:case}, our proposed model \modelname{1} predicted 10 true binding residues and 4 false positive residues, achieving a Rec of 0.64, a Pre of 0.53, and an F1 score of 0.58. In contrast, GraphSite predicted 8 true binding residues and 13 false positive residues, achieving a Rec of 0.60, a Pre of 0.47, and an F1 score of 0.52. On the other hand, GraphBind predicted only 6 true binding residues and 5 false positive residues, achieving a Rec of 0.40, a Pre of 0.26, and an F1 score of 0.32. Although GraphSite also predicted enough true binding residues, it had a higher false positive rate, whereas \modelname{1} exhibited higher accuracy.

These results are reasonable because: $\text {(i)}$ \modelname{1} considers atom-level segmentation, capturing local geometric information within amino acids, while traditional methods only consider atom features and spatial information between amino acids. $\text {(ii)}$ The LLM model characterizes the semantic information between amino acids better than models that use only local biological features. $\text {(iii)}$ \modelname{1} integrates sequence and structure representations on residue prediction task and filters interacting surface residues based on RPF, which is more in line with biological experimental logic. Our results are also a significant improvement over other methods on the more challenging DNA\_181 and RNA\_117 tests, demonstrating the great advantages of \modelname{1}.

\begin{table}[t]
\centering
\small
\begin{tabular}{lcccccr}
\hline
  Module &  F1 & MCC & AUC  & AUPR  \\
  \hline
  ALS$^{a}$  & 0.429 & 0.501 & 0.902  & 0.477 \\
  LLM$^{b}$  & 0.511 & 0.524 & 0.928  & 0.524 \\
  ALS$^{a}$ $+$ LLM$^{b}$ & 0.574 & 0.560 & 0.943  & 0.575\\
 \hline
 {\bf \modelname{1}} & {\bf 0.602} & {\bf 0.581} & {\bf 0.953}  & {\bf 0.628} \\
 - AWA$^{c}$ & 0.582 & 0.564 & 0.945  & 0.581\\
 - SSL$^{d}$ & 0.588 & 0.568 & 0.951  & 0.606\\
 - RPF$^{e}$  & 0.595 & 0.575 & 0.951  & 0.620\\
 - COE$^{f}$  & 0.599 & 0.577 & 0.952  & 0.624 \\ 
  \hline
  GraphBind & 0.512 & 0.484 & 0.912 & 0.497 \\ 
  GraphSite & 0.543 & 0.519 & 0.934 & 0.544 \\ 
\hline
\end{tabular}
\caption{Ablation study on DNA\_129 testset.}
\label{table:ablation}

\end{table}

\subsection{Ablation study}
\label{subsec:Ablation}
{\noindent\bf Modules incremental ablation.} We present several incremental ablation studies on the DNA\_129 dataset to evaluate the effectiveness of our proposed modules. As shown in the table~\ref{table:ablation}, using only the protein language model did not yield good results on the task. Similarly, using only the atom point cloud segmentation encoder led to even worse results. However, when combining the sequence and structure features, the results outperformed the state-of-the-art method by 5.7\% in AUPRC and 7.8\% in MCC. This demonstrated the effectiveness of our proposed cross-modal module, as the pre-trained language model contained no spatial information. The AWA module, which incorporates local geometric knowledge from atoms to amino acids, also contributed to the performance improvement. Removing this module led to significantly lower results. The SSL module improved the segmentation encoder's robustness on imbalanced data, and removing it resulted in a 2.2\% reduction in AUPR. Finally, the RPF module optimized the results of cross-modal predictions based on a priori knowledge of biology. Introducing conventional co-evolutionary features led to an insignificant performance drop, as the pre-trained language model already contained most of the feature information. Overall, these ablation studies confirmed the effectiveness of each proposed module in our \modelname{1} method.

{\noindent\bf Large language model (LLM) ablation.} 
Based on Table~\ref{table:LLM}, it can be observed that using a smaller LLM model with fewer transformer layers leads to lower performance on the task. Specifically, using only 12-layers of transformer leads to a 2.4\% decrease in AUC compared to 33-layers. Additionly, fine-tuning the LLM model on the task also greatly improves the performance. For example, fine-tuning the 33-layer LLM model leads to a 3.1\% improvement in AUPRC compared to using the pre-trained LLM model only. These results suggest that a larger and fine-tuned LLM model can better capture the language information in protein sequences, which is beneficial for the residues prediction.


{\noindent\bf Atom-wise attention.} We adopt a simple MLP-based attention to incorporate the fine-grained local geometric knowledge between atoms and amino acids. As shown in the table~\ref{table:Attention}, only average or stacking the atoms feature leads to a significant reduction on performance, which lost the local geometric knowledge. We replace the atom-wise attention with a single self-attention layer on cross-modal module, the result is less than stacking all atoms feature, probably because self-attention layer over-smooth the local geometric knowledge between all atoms.

\begin{table}[t]
\centering
\footnotesize
\begin{tabular}{lcccccr}
\hline
  Layers&Fine-Tune & EMB\_D  & MCC &AUC & AUPR \\
  \hline
 6 & N & 320 &0.432 & 0.904& 0.474\\
  6 & Y & 320 &0.439& 0.915 &0.487 \\
    \hline
 12 & N & 480 &0.457 & 0.920 &0.509 \\
 12 &Y & 480 &  0.492 & 0.929& 0.524 \\
 \hline
 30 &N & 640 & 0.525 &0.931 &0.547 \\
 30 &Y & 640 & 0.548 & 0.944& 0.589\\  
 \hline
 33 &N & 1280 & 0.559 &0.947 & 0.597 \\
 {\bf 33} &{\bf Y} & {\bf 1280} & {\bf 0.581} & {\bf 0.953} & {\bf 0.628}\\  
\hline
\end{tabular}
\caption{Ablations for pre-trained Large Language Model (ESM2) on DNA\_129 test. EMB\_D is the output dimension of the LLM.}
\label{table:LLM}
\end{table}

\begin{table}[t]
\centering
\footnotesize
\begin{tabular}{l|cr}
\hline
Attention Method &AUC & AUPR \\
  \hline
 Atom feature (Mean) & 0.931 & 0.568 \\
 Atom feature (Stack) & 0.945 & 0.583 \\
 Self-attention & 0.941 & 0.577 \\ 
 {\bf Atom-wise attention } & {\bf 0.953} & {\bf 0.628}\\
\hline
\end{tabular}
\caption{Ablations for Atom-wise attention module.}
\label{table:Attention}
\end{table}




\section{Conclusion}
\label{sec:conclusion}
In this study, we propose \modelname{1}, a cross-modal framework for identifying protein nucleic-acid-binding residues by using both protein structure and sequence information. In addition, we introduce an atom-wise attention module that captures the positional relationship between atoms and residues for extracting fine-grained local geometric representations to encode the 3D protein structures. Our method achieves state-of-the-art results on three benchmark datasets and outperforms other single-mode methods based on a comprehensive evaluation. 

In future work, we plan to further improve our structural encoder to extend various downstream tasks. This study provides evidence that cross-modal strategies are effective in protein-related tasks. Additionally, similar to the large language models based on protein sequence, a general 3D structure pre-training model also warrants further research. Besides, another tread is to solve the condition without native protein structure or reliable folding protein structures.

\section{Acknowledgment}
This work is partially supported by the National Key R\&D Program of China (NO.2022ZD0160101), by Shenzhen-Hong Kong Joint Funding No.SGDX20211123112401002, and by Shenzhen General Program No. JCYJ20220530143600001. This work is supported by funds from the Focus Project of AI for Science of Comprehensive Prosperity Plan for Disciplines of Fudan University, Netmind.AI, and Protagolabs Inc (to S.S.).

\bibliography{aaai24}

\begin{thebibliography}{42}
\providecommand{\natexlab}[1]{#1}

\bibitem[{Altschul et~al.(1997)Altschul, Madden, Sch{\"a}ffer, Zhang, Zhang, Miller, and Lipman}]{altschul1997gapped}
Altschul, S.~F.; Madden, T.~L.; Sch{\"a}ffer, A.~A.; Zhang, J.; Zhang, Z.; Miller, W.; and Lipman, D.~J. 1997.
\newblock Gapped BLAST and PSI-BLAST: a new generation of protein database search programs.
\newblock \emph{Nucleic acids research}, 25(17): 3389--3402.

\bibitem[{Baek et~al.(2021)Baek, DiMaio, Anishchenko, Dauparas, Ovchinnikov, Lee, Wang, Cong, Kinch, Schaeffer et~al.}]{baek2021accurate}
Baek, M.; DiMaio, F.; Anishchenko, I.; Dauparas, J.; Ovchinnikov, S.; Lee, G.~R.; Wang, J.; Cong, Q.; Kinch, L.~N.; Schaeffer, R.~D.; et~al. 2021.
\newblock Accurate prediction of protein structures and interactions using a three-track neural network.
\newblock \emph{Science}, 373(6557): 871--876.

\bibitem[{Bheemireddy et~al.(2022)Bheemireddy, Sandhya, Srinivasan, and Sowdhamini}]{bheemireddy2022computational}
Bheemireddy, S.; Sandhya, S.; Srinivasan, N.; and Sowdhamini, R. 2022.
\newblock Computational tools to study RNA-protein complexes.
\newblock \emph{Frontiers in Molecular Biosciences}, 9.

\bibitem[{Chen and Ludtke(2021)}]{chen2021deep}
Chen, M.; and Ludtke, S.~J. 2021.
\newblock Deep learning-based mixed-dimensional Gaussian mixture model for characterizing variability in cryo-EM.
\newblock \emph{Nature methods}, 18(8): 930--936.

\bibitem[{Chen et~al.(2022)Chen, Zhou, Wang, Wang, He, Hu, Ding, Guan, and He}]{chen2022part}
Chen, T.; Zhou, D.; Wang, J.; Wang, S.; He, Q.; Hu, C.; Ding, E.; Guan, Y.; and He, X. 2022.
\newblock Part-aware Prototypical Graph Network for One-shot Skeleton-based Action Recognition.
\newblock \emph{arXiv preprint arXiv:2208.09150}.

\bibitem[{Chen et~al.(2020)Chen, Fan, Girshick, and He}]{chen2020improved}
Chen, X.; Fan, H.; Girshick, R.; and He, K. 2020.
\newblock Improved baselines with momentum contrastive learning.
\newblock \emph{arXiv preprint arXiv:2003.04297}.

\bibitem[{Chen and Lim(2008)}]{chen2008predicting}
Chen, Y.~C.; and Lim, C. 2008.
\newblock Predicting RNA-binding sites from the protein structure based on electrostatics, evolution and geometry.
\newblock \emph{Nucleic acids research}, 36(5): e29.

\bibitem[{Chou(2001)}]{chou2001prediction}
Chou, K.-C. 2001.
\newblock Prediction of protein cellular attributes using pseudo-amino acid composition.
\newblock \emph{Proteins: Structure, Function, and Bioinformatics}, 43(3): 246--255.

\bibitem[{Dai and Bailey-Kellogg(2021)}]{dai2021protein}
Dai, B.; and Bailey-Kellogg, C. 2021.
\newblock Protein interaction interface region prediction by geometric deep learning.
\newblock \emph{Bioinformatics}, 37(17): 2580--2588.

\bibitem[{Ding et~al.(2021)Ding, Liu, Wang, and Jiang}]{ding2021vision}
Ding, H.; Liu, C.; Wang, S.; and Jiang, X. 2021.
\newblock Vision-language transformer and query generation for referring segmentation.
\newblock In \emph{Proceedings of the IEEE/CVF International Conference on Computer Vision}, 16321--16330.

\bibitem[{Gr{\o}nning et~al.(2020)Gr{\o}nning, Doktor, Larsen, Petersen, Holm, Bruun, Hansen, Hartung, Baumbach, and Andresen}]{gronning2020deepclip}
Gr{\o}nning, A. G.~B.; Doktor, T.~K.; Larsen, S.~J.; Petersen, U. S.~S.; Holm, L.~L.; Bruun, G.~H.; Hansen, M.~B.; Hartung, A.-M.; Baumbach, J.; and Andresen, B.~S. 2020.
\newblock DeepCLIP: predicting the effect of mutations on protein--RNA binding with deep learning.
\newblock \emph{Nucleic acids research}, 48(13): 7099--7118.

\bibitem[{Guo et~al.(2020)Guo, Wang, Hu, Liu, Liu, and Bennamoun}]{guo2020deep}
Guo, Y.; Wang, H.; Hu, Q.; Liu, H.; Liu, L.; and Bennamoun, M. 2020.
\newblock Deep learning for 3d point clouds: A survey.
\newblock \emph{IEEE transactions on pattern analysis and machine intelligence}, 43(12): 4338--4364.

\bibitem[{Jumper et~al.(2021)Jumper, Evans, Pritzel, Green, Figurnov, Ronneberger, Tunyasuvunakool, Bates, {\v{Z}}{\'\i}dek, Potapenko et~al.}]{jumper2021highly}
Jumper, J.; Evans, R.; Pritzel, A.; Green, T.; Figurnov, M.; Ronneberger, O.; Tunyasuvunakool, K.; Bates, R.; {\v{Z}}{\'\i}dek, A.; Potapenko, A.; et~al. 2021.
\newblock Highly accurate protein structure prediction with AlphaFold.
\newblock \emph{Nature}, 596(7873): 583--589.

\bibitem[{Kim, Yura, and Go(2006)}]{kim2006amino}
Kim, O.~T.; Yura, K.; and Go, N. 2006.
\newblock Amino acid residue doublet propensity in the protein--RNA interface and its application to RNA interface prediction.
\newblock \emph{Nucleic acids research}, 34(22): 6450--6460.

\bibitem[{Lam et~al.(2019)Lam, Li, Zhu, Umarov, Jiang, H{\'e}liou, Sheong, Liu, Long, Li et~al.}]{lam2019deep}
Lam, J.~H.; Li, Y.; Zhu, L.; Umarov, R.; Jiang, H.; H{\'e}liou, A.; Sheong, F.~K.; Liu, T.; Long, Y.; Li, Y.; et~al. 2019.
\newblock A deep learning framework to predict binding preference of RNA constituents on protein surface.
\newblock \emph{Nature communications}, 10(1): 4941.

\bibitem[{Li et~al.(2014)Li, Yamashita, Amada, and Standley}]{li2014quantifying}
Li, S.; Yamashita, K.; Amada, K.~M.; and Standley, D.~M. 2014.
\newblock Quantifying sequence and structural features of protein--RNA interactions.
\newblock \emph{Nucleic acids research}, 42(15): 10086--10098.

\bibitem[{Lin et~al.(2022)Lin, Akin, Rao, Hie, Zhu, Lu, Smetanin, Verkuil, Kabeli, Shmueli et~al.}]{lin2022evolutionary}
Lin, Z.; Akin, H.; Rao, R.; Hie, B.; Zhu, Z.; Lu, W.; Smetanin, N.; Verkuil, R.; Kabeli, O.; Shmueli, Y.; et~al. 2022.
\newblock Evolutionary-scale prediction of atomic level protein structure with a language model.
\newblock \emph{bioRxiv}, 2022--07.

\bibitem[{Liu et~al.(2021)Liu, HaoChen, Gaidon, and Ma}]{liu2021self}
Liu, H.; HaoChen, J.~Z.; Gaidon, A.; and Ma, T. 2021.
\newblock Self-supervised learning is more robust to dataset imbalance.
\newblock \emph{arXiv preprint arXiv:2110.05025}.

\bibitem[{Liu and Hu(2013)}]{liu2013dnabind}
Liu, R.; and Hu, J. 2013.
\newblock DNABind: A hybrid algorithm for structure-based prediction of DNA-binding residues by combining machine learning-and template-based approaches.
\newblock \emph{PROTEINS: structure, Function, and Bioinformatics}, 81(11): 1885--1899.

\bibitem[{McGinnis and Madden(2004)}]{mcginnis2004blast}
McGinnis, S.; and Madden, T.~L. 2004.
\newblock BLAST: at the core of a powerful and diverse set of sequence analysis tools.
\newblock \emph{Nucleic acids research}, 32(suppl\_2): W20--W25.

\bibitem[{Pan and Shen(2018)}]{pan2018predicting}
Pan, X.; and Shen, H.-B. 2018.
\newblock Predicting RNA--protein binding sites and motifs through combining local and global deep convolutional neural networks.
\newblock \emph{Bioinformatics}, 34(20): 3427--3436.

\bibitem[{Panda et~al.(2021)Panda, Chen, Fan, Sun, Saenko, Oliva, and Feris}]{panda2021adamml}
Panda, R.; Chen, C.-F.~R.; Fan, Q.; Sun, X.; Saenko, K.; Oliva, A.; and Feris, R. 2021.
\newblock Adamml: Adaptive multi-modal learning for efficient video recognition.
\newblock In \emph{Proceedings of the IEEE/CVF International Conference on Computer Vision}, 7576--7585.

\bibitem[{Remmert et~al.(2012)Remmert, Biegert, Hauser, and S{\"o}ding}]{remmert2012hhblits}
Remmert, M.; Biegert, A.; Hauser, A.; and S{\"o}ding, J. 2012.
\newblock HHblits: lightning-fast iterative protein sequence searching by HMM-HMM alignment.
\newblock \emph{Nature methods}, 9(2): 173--175.

\bibitem[{Saito and Rehmsmeier(2015)}]{saito2015precision}
Saito, T.; and Rehmsmeier, M. 2015.
\newblock The precision-recall plot is more informative than the ROC plot when evaluating binary classifiers on imbalanced datasets.
\newblock \emph{PloS one}, 10(3): e0118432.

\bibitem[{Schmohl and S{\"o}rgel(2019)}]{schmohl2019submanifold}
Schmohl, S.; and S{\"o}rgel, U. 2019.
\newblock Submanifold sparse convolutional networks for semantic segmentation of large-scale ALS point clouds.
\newblock \emph{ISPRS Annals of the Photogrammetry, Remote Sensing and Spatial Information Sciences}, 4: 77--84.

\bibitem[{Su et~al.(2019)Su, Liu, Sun, Peng, and Yang}]{su2019improving}
Su, H.; Liu, M.; Sun, S.; Peng, Z.; and Yang, J. 2019.
\newblock Improving the prediction of protein--nucleic acids binding residues via multiple sequence profiles and the consensus of complementary methods.
\newblock \emph{Bioinformatics}, 35(6): 930--936.

\bibitem[{Sverrisson et~al.(2021)Sverrisson, Feydy, Correia, and Bronstein}]{sverrisson2021fast}
Sverrisson, F.; Feydy, J.; Correia, B.~E.; and Bronstein, M.~M. 2021.
\newblock Fast end-to-end learning on protein surfaces.
\newblock In \emph{Proceedings of the IEEE/CVF Conference on Computer Vision and Pattern Recognition}, 15272--15281.

\bibitem[{Verkuil et~al.(2022)Verkuil, Kabeli, Du, Wicky, Milles, Dauparas, Baker, Ovchinnikov, Sercu, and Rives}]{verkuil2022language}
Verkuil, R.; Kabeli, O.; Du, Y.; Wicky, B.~I.; Milles, L.~F.; Dauparas, J.; Baker, D.; Ovchinnikov, S.; Sercu, T.; and Rives, A. 2022.
\newblock Language models generalize beyond natural proteins.
\newblock \emph{bioRxiv}, 2022--12.

\bibitem[{Wang et~al.(2022)Wang, Wu, Duan, and Huang}]{wang2022point}
Wang, Y.; Wu, S.; Duan, Y.; and Huang, Y. 2022.
\newblock A point cloud-based deep learning strategy for protein--ligand binding affinity prediction.
\newblock \emph{Briefings in Bioinformatics}, 23(1): bbab474.

\bibitem[{Wei et~al.(2022)Wei, Chen, Zong, Gao, and Li}]{wei2022protein}
Wei, J.; Chen, S.; Zong, L.; Gao, X.; and Li, Y. 2022.
\newblock Protein--RNA interaction prediction with deep learning: structure matters.
\newblock \emph{Briefings in bioinformatics}, 23(1): bbab540.

\bibitem[{Wu et~al.(2018)Wu, Peng, Zhang, and Yang}]{wu2018coach}
Wu, Q.; Peng, Z.; Zhang, Y.; and Yang, J. 2018.
\newblock COACH-D: improved protein--ligand binding sites prediction with refined ligand-binding poses through molecular docking.
\newblock \emph{Nucleic acids research}, 46(W1): W438--W442.

\bibitem[{Xia et~al.(2021)Xia, Xia, Pan, and Shen}]{xia2021graphbind}
Xia, Y.; Xia, C.-Q.; Pan, X.; and Shen, H.-B. 2021.
\newblock GraphBind: protein structural context embedded rules learned by hierarchical graph neural networks for recognizing nucleic-acid-binding residues.
\newblock \emph{Nucleic acids research}, 49(9): e51--e51.

\bibitem[{Xue, Dobbs, and Honavar(2011)}]{xue2011homppi}
Xue, L.~C.; Dobbs, D.; and Honavar, V. 2011.
\newblock HomPPI: a class of sequence homology based protein-protein interface prediction methods.
\newblock \emph{BMC bioinformatics}, 12(1): 1--24.

\bibitem[{Yan, Friedrich, and Kurgan(2016)}]{yan2016comprehensive}
Yan, J.; Friedrich, S.; and Kurgan, L. 2016.
\newblock A comprehensive comparative review of sequence-based predictors of DNA-and RNA-binding residues.
\newblock \emph{Briefings in bioinformatics}, 17(1): 88--105.

\bibitem[{Yan et~al.(2022)Yan, Lu, Li, Wei, Gao, Wang, Wu, and Cui}]{yan2022pointsite}
Yan, X.; Lu, Y.; Li, Z.; Wei, Q.; Gao, X.; Wang, S.; Wu, S.; and Cui, S. 2022.
\newblock PointSite: a point cloud segmentation tool for identification of protein ligand binding atoms.
\newblock \emph{Journal of Chemical Information and Modeling}, 62(11): 2835--2845.

\bibitem[{Yang, Roy, and Zhang(2012)}]{yang2012biolip}
Yang, J.; Roy, A.; and Zhang, Y. 2012.
\newblock BioLiP: a semi-manually curated database for biologically relevant ligand--protein interactions.
\newblock \emph{Nucleic acids research}, 41(D1): D1096--D1103.

\bibitem[{Yu et~al.(2013)Yu, Hu, Yang, Shen, Tang, and Yang}]{yu2013designing}
Yu, D.-J.; Hu, J.; Yang, J.; Shen, H.-B.; Tang, J.; and Yang, J.-Y. 2013.
\newblock Designing template-free predictor for targeting protein-ligand binding sites with classifier ensemble and spatial clustering.
\newblock \emph{IEEE/ACM transactions on computational biology and bioinformatics}, 10(4): 994--1008.

\bibitem[{Yuan et~al.(2022)Yuan, Chen, Rao, Zheng, Zhao, and Yang}]{yuan2022alphafold2}
Yuan, Q.; Chen, S.; Rao, J.; Zheng, S.; Zhao, H.; and Yang, Y. 2022.
\newblock AlphaFold2-aware protein--DNA binding site prediction using graph transformer.
\newblock \emph{Briefings in Bioinformatics}, 23(2): bbab564.

\bibitem[{Zhang, Chen, and Liu(2021)}]{zhang2021ncbrpred}
Zhang, J.; Chen, Q.; and Liu, B. 2021.
\newblock NCBRPred: predicting nucleic acid binding residues in proteins based on multilabel learning.
\newblock \emph{Briefings in bioinformatics}, 22(5): bbaa397.

\bibitem[{Zhang and Skolnick(2005)}]{zhang2005tm}
Zhang, Y.; and Skolnick, J. 2005.
\newblock TM-align: a protein structure alignment algorithm based on the TM-score.
\newblock \emph{Nucleic acids research}, 33(7): 2302--2309.

\bibitem[{Zhou, Ruan, and Canu(2019)}]{zhou2019review}
Zhou, T.; Ruan, S.; and Canu, S. 2019.
\newblock A review: Deep learning for medical image segmentation using multi-modality fusion.
\newblock \emph{Array}, 3: 100004.

\bibitem[{Zhu et~al.(2019)Zhu, Hu, Song, and Yu}]{zhu2019dnapred}
Zhu, Y.-H.; Hu, J.; Song, X.-N.; and Yu, D.-J. 2019.
\newblock DNAPred: accurate identification of DNA-binding sites from protein sequence by ensembled hyperplane-distance-based support vector machines.
\newblock \emph{Journal of chemical information and modeling}, 59(6): 3057--3071.

\end{thebibliography}

\newpage

\section{Supplementary Material}

\section{Evalution metrics}
\label{sec:evalution}
We utilized similar evaluation metrics to previous studies~\cite{xia2021graphbind,yuan2022alphafold2}, including precision (Pre), recall (Rec), F1-score (F1), Matthews correlation coefficient (MCC), area under the receiver operating characteristic curve (AUC), and area under the precision-recall curve (AUPR), to assess the predictive performance of our model.

\begin{equation}
Pre =\frac{\mathrm{TP}}{\mathrm{TP}+\mathrm{FP}}
\end{equation}

\begin{equation}
Rec =\frac{\mathrm{TP}}{\mathrm{TP}+\mathrm{FN}}
\end{equation}

\begin{equation}
\mathrm{F} 1=2 \times \frac{\text { Precision } \times \text { Recall }}{\text { Precision }+ \text { Recall }}
\label{eq:f1}
\end{equation}

\begin{footnotesize}
\begin{align} 
\mathrm{MCC}=\frac{\mathrm{TP} \times \mathrm{TN}-\mathrm{FN} \times \mathrm{FP}}{\sqrt{(\mathrm{TP}+\mathrm{FP}) \times(\mathrm{TP}+\mathrm{FN}) \times(\mathrm{TN}+\mathrm{FP}) \times(\mathrm{TN}+\mathrm{FN})}}
\end{align}
\end{footnotesize}


The evaluation metrics used in this study included true positives (TP) and true negatives (TN) to measure the number of correctly identified binding and non-binding sites, respectively, as well as false positives (FP) and false negatives (FN) to represent the number of incorrect predictions. In addition, we used area under the receiver operating characteristic curve (AUC) and area under the precision-recall curve (AUPR) to evaluate the overall performance of the model, as they are independent of thresholds. The other metrics, including precision (Pre), recall (Rec), and F1-score (F1), were calculated using a threshold to convert predicted binding probabilities to binary predictions. The threshold was determined by maximizing the F1-score for the model. For hyperparameter selection, we used AUPR because it is more sensitive and emphasizes the minority class in imbalanced two-class classification tasks~\cite{saito2015precision}.

\section{Notations}
\label{sec:notation}
We summarize the notations used in the paper in Table~\ref{table:notion}. The notations are listed under 4 groups: Subscript, Network, Representation and Operation.

  


\begin{table}[!t]
\caption{Notation Table.}
\begin{center}
\resizebox{\linewidth}{!}{
\begin{tabular}{ccc}
\hline
  & Symbol  & Description  \\
  \hline
  
Subscript & $i$ & Amino acid index in a protein \\ 
 & $k$ & The maximum number of atoms \\
 & $j$ & Atom index in an amino acid \\
 & $K$ &Maximum atom number limit \\
Network & $F_\Theta$ & Sparse convolution encoder\\
& $G_\psi$ &  MLP projection head \\
Representation& $a_{ij}$ & Atom representation \\
& $A_i$ & Amino acid representation\\
& $F_{struct}$ &  Protein structural representation \\ 
& $F_{seq}$ &  Protein sequence representation \\ 
& $Q^t$ &  Input data augmentation version \\
& $z$ & Projected vectors \\ 
& $\xi$ & Amino acid propensity  \\
Operation& $\Lambda$ & Atom-wise attention score list\\ 
& $[.]$ & The concatenate function \\ 
& $\odot$  & The element-wise multiplication \\
& $+$ &  The element-wise addition \\ 

\hline
\end{tabular}
}
\end{center}
\label{table:notion}
\end{table}

\section{Conventional co-evolutionary features}

Co-evolving amino acids can interact with each other at the structural level, and evolutionarily conserved residues may contain motifs that are important for protein properties such as DNA-binding propensity. In this study, we examined two types of co-evolutionary features: (i) position-specific scoring matrix (PSSM), and (ii) hidden Markov model (HMM) profiles.

To generate the PSSM, we used PSI-BLAST~\cite{altschul1997gapped} to search the query sequence against the UniRef90 database with an E-value of 0.001 and three iterations. For the HMM profile, we used HHblits~\cite{remmert2012hhblits} to align the query sequence against the UniClust30 database with default parameters. Each amino acid was encoded as a 20-dimensional vector in both the PSSM and HMM profiles. We normalized the values to scores between 0 and 1 using Equation~\ref{eq:rec}:

\begin{equation}
\mathrm{U}_{\text {norm }}=\frac{\mathrm{U}-\operatorname{U_{Min}}}{\operatorname{U_{Max}}-\operatorname{U_{Min}}}
\label{eq:rec}
\end{equation}

where $U$ is the original feature value, and Min and Max are the smallest and biggest values of this feature type observed in the training set.

\section{Algorithmic Specifications of the RBF}
Algorithm~\ref{alg:algorithm} details the Residue propensity filter (RBF) module.

\begin{algorithm}[t]
\caption{Residue Propensity Filter (RPF) Algorithm}\label{nab}
\label{alg:algorithm}
\begin{algorithmic}
\renewcommand{\algorithmicrequire}{\textbf{Input:}}
\renewcommand{\algorithmicensure}{\textbf{Output:}}
\REQUIRE{Amino acid sequence $AAs$\\ 
\qquad \ Predicted logits $logits$ \\ 
\qquad \ Positive propensity threshold $\alpha$\\
\qquad \ Negative propensity threshold $\beta$}
\ENSURE{Filtered logits $logits$}
\FOR{$i \gets 0$ to $|AAs|$}
\STATE $K_i \gets$ k-nearest amino acids to $a_i$ using surface distance
\IF{$|\{a \in K_i : logits_a > 0.5\}| > round(0.8*k)$ and $propensity_{a_i} > \alpha$}
\STATE $logits_{a_i} \gets logits_{a_i} * 1.2$
\ELSIF{$|\{a \in K_i : logits_a < 0.5\}| > round(0.8*k)$ and $propensity_{a_i} < \beta$}
\STATE $logits_{a_i} \gets logits_{a_i} * 0.8$
\ENDIF
\ENDFOR
\end{algorithmic}
\end{algorithm}
\vspace{-6mm}

\begin{figure}[!t]
\begin{center}
\includegraphics[width=0.9\linewidth]{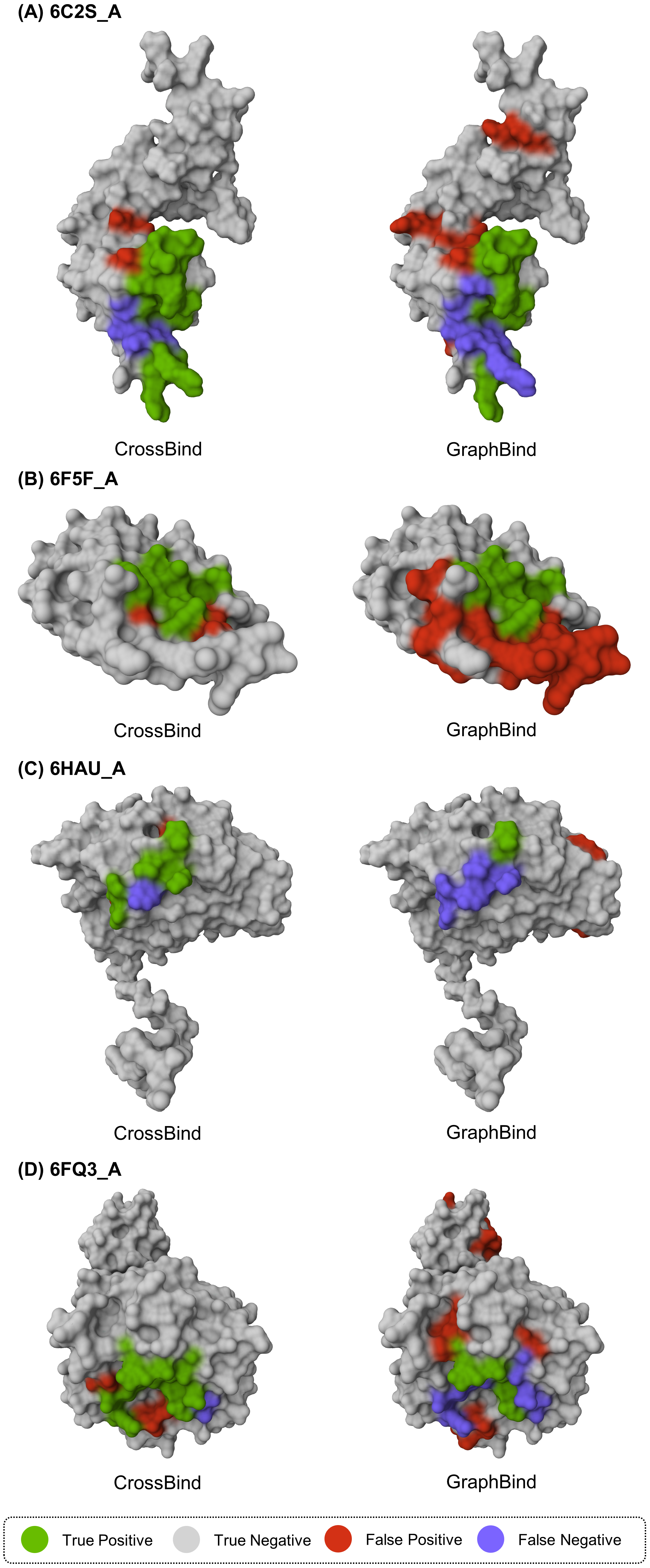}
\end{center}
   \caption{Visualization of four cases predicted by \modelname{1} and GraphBind. (A), (B) are from DNA\_129 and (C), (D) are from RNA\_117.}
\label{fig:all}
\end{figure}

\section{Ablation Study}[t]
\label{sec:abl}
Table~\ref{table:2} presents the results of the complete module ablation experiment on three test datasets: DNA\_129, DNA\_181, and RNA\_117. Our findings indicate that using only the structural or sequence information of a protein leads to poor results in this task. Specifically, when using only structural information with an atom-level segmentation (ALS) encoder, the F1 score is reduced by 40.3\%, 48.9\%, and 49.4\% on DNA\_129, DNA\_181, and RNA\_117, respectively. Similarly, when using only sequence information with a large language model (LLM), the F1 score is reduced by 17.5\%, 18.4\%, and 14.7\% on DNA\_129, DNA\_181, and RNA\_117, respectively.

Although the LLM performs better than the ALS due to the large number of pre-training protein samples and parameters, it still falls far below the performance of \modelname{1}, demonstrating the effectiveness of our proposed cross-modal learning strategy. Specifically, we propose several optimization modules based on the identification of nucleic acid-binding residues. Among these modules, the atom-wise attention (AWA) module provides the greatest benefit, and its removal results in a decrease of the F1 score by 3.4\%, 3.2\%, and 6.3\% on DNA\_129, DNA\_181, and RNA\_117, respectively.

\begin{table*}[!t]
\begin{small}
\caption{Ablation study.}
\label{table:2}
\begin{center}
\begin{tabular*}{1\textwidth}{@{\extracolsep{\fill}}ccccccc|ccccc} 
\hline

 Dataset & ALS$^{a}$ & LLM$^{b}$ & AWA$^{c}$ & SSL$^{d}$ & COE$^{e}$ & RPF$^{f}$ & F1 & MCC & AUC  & AUPR  \\
\hline
DNA\_{129} & $\checkmark$ & $\checkmark$ & $\checkmark$ & $\checkmark$ & $\checkmark$ & $\checkmark$ & {\bf 0.602} & {\bf 0.581} & {\bf 0.953}  & {\bf 0.628} \\
 & $\checkmark$  & $\checkmark$ & $\checkmark$ & $\checkmark$ & $\checkmark$& $\times$  & 0.595 & 0.575 & 0.951  & 0.620 \\
&$\checkmark$ & $\checkmark$& $\checkmark$ & $\checkmark$ & $\times$& $\checkmark$  & 0.599 & 0.577 & 0.952  & 0.624 \\
&$\checkmark$  & $\checkmark$ &  $\checkmark$& $\times$ &$\checkmark$ & $\checkmark$ & 0.588 & 0.568 & 0.951  & 0.606 \\
&$\checkmark$  & $\checkmark$ &$\times$ & $\checkmark$ &$\checkmark$ &$\checkmark$& 0.582 & 0.564 & 0.945  & 0.581 \\
 & $\checkmark$ & $\checkmark$ & $\times$ &$\times$&$\times$ &$\times$ & 0.574 & 0.560 & 0.943  & 0.575 \\
 &$\checkmark$  &  $\times$& $\checkmark$  & $\checkmark$ &$\checkmark$ & $\checkmark$  & 0.486 & 0.502 & 0.919  & 0.527 \\
 & $\times$ & $\checkmark$ & $\times$ & $\times$ &$\checkmark$ & $\checkmark$  & 0.515 & 0.535 & 0.929  & 0.535 \\
 &     $\checkmark$ &$\times$  &  $\times$ &  $\times$&$\times$ & $\times$ & 0.429 & 0.501 & 0.902  & 0.477 \\  
  &   $\checkmark$ & $\times$ & $\checkmark$  & $\times$ & $\times$& $\times$ & 0.435 & 0.508 & 0.908  & 0.485 \\    
   &     $\checkmark$ & $\times$ &  $\times$&$\checkmark$  & $\times$& $\times$ & 0.433 & 0.506 & 0.905  & 0.490 \\   
  & $\checkmark$ &  $\times$& $\checkmark$  & $\checkmark$ & $\times$& $\times$ & 0.445 & 0.511 & 0.910  & 0.495 \\            
& $\times$  &$\checkmark$  & $\times$ & $\times$&$\times$ &  $\times$ & 0.511 & 0.524 & 0.928  & 0.524 \\
&  $\times$ & $\times$ & $\times$ &$\times$ & $\checkmark$ & $\times$  & 0.382 & 0.334 & 0.785  & 0.358 \\
\hline

RNA\_{181} & $\checkmark$ & $\checkmark$ & $\checkmark$ & $\checkmark$ & $\checkmark$ & $\checkmark$ & {\bf 0.475} & {\bf 0.448} & {\bf 0.932}  & {\bf 0.424} \\
 & $\checkmark$  & $\checkmark$ & $\checkmark$ & $\checkmark$ & $\checkmark$& $\times$  & 0.469 & 0.443 & 0.930  & 0.418 \\
& $\checkmark$ & $\checkmark$& $\checkmark$ & $\checkmark$ & $\times$& $\checkmark$  & 0.473 & 0.445 & 0.931  & 0.420 \\
& $\checkmark$  & $\checkmark$ &  $\checkmark$& $\times$ &$\checkmark$ & $\checkmark$ & 0.464 & 0.435 & 0.929  & 0.402 \\
& $\checkmark$  & $\checkmark$ &$\times$ & $\checkmark$ &$\checkmark$ &$\checkmark$& 0.460 & 0.431 & 0.924  & 0.379 \\
&   $\checkmark$ & $\checkmark$ & $\times$ &$\times$&$\times$ &$\times$ & 0.451 & 0.426 & 0.921  & 0.372 \\
 & $\checkmark$  &  $\times$& $\checkmark$  & $\checkmark$ &$\checkmark$ & $\checkmark$  & 0.361 & 0.466 & 0.891  & 0.331 \\
 & $\times$ & $\checkmark$ & $\times$ & $\times$ &$\checkmark$ & $\checkmark$  & 0.392 & 0.492 & 0.910  & 0.342 \\
   &   $\checkmark$ &$\times$  &  $\times$ &  $\times$&$\times$ & $\times$ & 0.319 & 0.466 & 0.875  & 0.278 \\  
 &    $\checkmark$ & $\times$ & $\checkmark$  & $\times$ & $\times$& $\times$ & 0.326 & 0.472 & 0.881  & 0.298 \\    
  &      $\checkmark$ & $\times$ &  $\times$&$\checkmark$  & $\times$& $\times$ & 0.322 & 0.469 & 0.886  & 0.295 \\   
 &  $\checkmark$ &  $\times$& $\checkmark$  & $\checkmark$ & $\times$& $\times$ & 0.334 & 0.481 & 0.898  & 0.306 \\            
& $\times$  &$\checkmark$  & $\times$ & $\times$&$\times$ &  $\times$ & 0.401 & 0.495 & 0.909  & 0.304 \\
 & $\times$ & $\times$ & $\times$ &$\times$ & $\checkmark$ & $\times$  & 0.281 & 0.254 & 0.699  & 0.148 \\
\hline
RNA\_{117} & $\checkmark$ & $\checkmark$ & $\checkmark$ & $\checkmark$ & $\checkmark$ & $\checkmark$ & {\bf 0.420} & {\bf 0.402} & {\bf 0.903}  & {\bf 0.352} \\
 & $\checkmark$  & $\checkmark$ & $\checkmark$ & $\checkmark$ & $\checkmark$& $\times$  & 0.414 & 0.395 & 0.900  & 0.342 \\
& $\checkmark$ & $\checkmark$& $\checkmark$ & $\checkmark$ & $\times$& $\checkmark$  & 0.416 & 0.398 & 0.902  & 0.348 \\
& $\checkmark$  & $\checkmark$ &  $\checkmark$& $\times$ &$\checkmark$ & $\checkmark$ & 0.407 & 0.389 & 0.901  & 0.328 \\
& $\checkmark$  & $\checkmark$ &$\times$ & $\checkmark$ &$\checkmark$ &$\checkmark$& 0.404 & 0.382 & 0.894  & 0.302 \\
&   $\checkmark$ & $\checkmark$ & $\times$ &$\times$&$\times$ &$\times$ & 0.395 & 0.334 & 0.891  & 0.290 \\
 & $\checkmark$  &  $\times$& $\checkmark$  & $\checkmark$ &$\checkmark$ & $\checkmark$  & 0.301 & 0.272 & 0.872  & 0.254 \\
 & $\times$ & $\checkmark$ & $\times$ & $\times$ &$\checkmark$ & $\checkmark$  & 0.334 & 0.301 & 0.896  & 0.280 \\
   &   $\checkmark$ &$\times$  &  $\times$ &  $\times$&$\times$ & $\times$ & 0.275 & 0.266 & 0.867  & 0.229 \\  
 &    $\checkmark$ & $\times$ & $\checkmark$  & $\times$ & $\times$& $\times$ & 0.294 & 0.281 & 0.874  & 0.243 \\    
  &      $\checkmark$ & $\times$ &  $\times$&$\checkmark$  & $\times$& $\times$ & 0.290 & 0.277 & 0.871  & 0.231 \\   
 &  $\checkmark$ &  $\times$& $\checkmark$  & $\checkmark$ & $\times$& $\times$ & 0.298 & 0.288 & 0.878  & 0.254 \\            
& $\times$  &$\checkmark$  & $\times$ & $\times$&$\times$ &  $\times$ & 0.366 & 0.311 & 0.894  & 0.274 \\
 & $\times$ & $\times$ & $\times$ &$\times$ & $\checkmark$ & $\times$  & 0.154 & 0.135 & 0.543  & 0.122 \\
 \hline
\end{tabular*}
\end{center}
\end{small}
\end{table*}

{\bf Module inference time.} We evaluated the efficiency of \modelname{1} by measuring the inference time and frames per second (FPS) of all modules containing neural networks. Given the varying lengths of the proteins in the DNA\_129, we computed the average values across all samples. The results show \modelname{1} achieves an FPS of approximately 10, which is 10 times faster than GraphBind. While GraphSite also employs a protein language encoder, AlphaFold2, which requires protein structure prediction before encoding, making it challenging to define their inference time.

\section{Selected Visualizations}
\label{sec:vis}
Figure~\ref{fig:all} presents randomly selected protein samples for binding residue prediction using our proposed \modelname{1} and the GraphBind~\cite{xia2021graphbind} model on both DNA (6C2S\_A, 6F5F\_A) and RNA (6HAU\_A, 6FQ3\_A) test sets. These examples demonstrate the efficacy of our model, which outperforms GraphBind in accurately predicting nucleic acid-binding residues

\end{document}